\documentclass[sigconf, authorversion]{acmart}

\usepackage{enumitem}
\usepackage{framed}

\newsavebox{\quotebox}

\newenvironment{sidequote}
{\par\vspace{9pt}\begin{lrbox}{\quotebox}\begin{minipage}{\dimexpr\linewidth-1em-1pt-6pt\relax}\raggedright\small}
{\end{minipage}\end{lrbox}%
\noindent\hspace{1em}\vrule height\ht\quotebox depth\dp\quotebox width 1pt\hspace{6pt}\usebox{\quotebox}\par\vspace{9pt}}

\AtBeginDocument{%
  }

\copyrightyear{2027}
\acmYear{2027}
\acmConference[SIGCSE-TS 2027]{ACM Special Interest Group on Computer Science Education (SIGCSE)}{February 17--20, 2027}{Sacramento, CA, USA}
\acmBooktitle{2027 Technical Symposium on Computer Science Education (SIGCSE TS), February 17--20, 2027, Sacramento, CA, USA}
\acmDOI{}
\begin{document}

\title[Testing Our Foundations: Hallucinations in the Computing Education Literature]{Testing Our Foundations: Citation Trends, Errors, and Emerging Hallucinations in the Computing Education Literature}
%\title[Testing Our Foundations]{Testing Our Foundations: Citation Trends, Errors, and Emerging Hallucinations in the Computing Education Literature}

\author{Paul Denny}
\orcid{0000-0002-5150-9806}
\affiliation{
  \institution{University of Auckland}
  \city{Auckland}
  \country{New Zealand}}
\email{paul@cs.auckland.ac.nz}

\author{Gweneth Barbre}
\orcid{0009-0007-3125-4168}
\affiliation{%
  \institution{Abilene Christian University}
  \city{Abilene}
  \state{Texas}
  \country{USA}}
\email{gab23c@acu.edu}

\author{Musa Blake}
\orcid{0009-0001-2457-9147}
\affiliation{%
  \institution{Abilene Christian University}
  \city{Abilene}
  \state{Texas}
  \country{USA}}
\email{mbb23c@acu.edu}

\author{Yan Cathy Hua}
\orcid{0000-0001-9155-9667}
\affiliation{
  \institution{University of Auckland}
  \city{Auckland}
  \country{New Zealand}}
\email{yhua219@aucklanduni.ac.nz}

\author{Juho Leinonen}
\orcid{0000-0001-6829-9449}
\affiliation{
  \institution{Aalto University}
  \city{Espoo}
  \country{Finland}}
\email{juho.2.leinonen@aalto.fi}

\author{Andrew Luxton-Reilly}
\orcid{0000-0001-8269-2909}
\affiliation{%
  \institution{University of Auckland}
  \streetaddress{Private Bag 92019}
  \city{Auckland}
  \country{New Zealand}
}
\email{a.luxton-reilly@auckland.ac.nz}

\author{James Prather}
\orcid{0000-0003-2807-6042}
\affiliation{
  \institution{Abilene Christian University}
  \city{Abilene}
  \state{TX}
  \country{USA}
}
\email{james.prather@acu.edu}

\author{Brent N. Reeves}
\orcid{0000-0001-5781-1136}
\affiliation{%
   \institution{Abilene Christian University}
   \city{Abilene}
   \state{TX}
   \country{USA}
}
\email{brent.reeves@acu.edu}

\renewcommand{\shortauthors}{Paul Denny et al.}


\begin{abstract}

Accurate references are foundational to scholarly work, enabling verification, attribution, and systematic review. However, the rapid adoption of large language models has introduced a serious integrity concern: plausible-looking but fabricated citations. Although hallucinated references are widely discussed, their visibility within specific research communities remains unclear. We address this gap by examining reference integrity at key computing education venues using ACM Digital Library data. We analyze referencing trends across 24,751 computing education papers and compare them with the broader ACM corpus of more than 723,000 papers and 15 million references. We then examine reference lists from these venues, classify common bibliographic errors, and manually identify LLM-generated hallucinations containing verifiably false information, including impossible page ranges, invented titles, and misattributed authors. In 2025, hallucinated references appeared across five SIGCSE-sponsored or in-cooperation venues. At the Technical Symposium alone, verified hallucinated references increased from 3 in 2025 to 17 in 2026, appearing in 2.3\% of 2026 proceedings papers. Although still relatively rare for now, this growth poses an integrity risk our community should not ignore.

\end{abstract}



% \ccsdesc[500]{Applied computing~Education}

%, Hallucinated citations, Reference integrity}

\maketitle

\section{Introduction}
References are an important foundation of scholarly work. They allow readers to verify claims, trace the development of ideas, credit prior work, and build cumulative knowledge. They also serve as data in citation analyses, systematic reviews, and scientometric studies.  Such studies serve an important role in helping the computing education community understand its own development, including the evolution of introductory programming research, and the broader thematic landscape of the field \cite{lister2008citation,becker2019fifty,papamitsiou2020computing,lopezpernas2023scientometric}.

\begin{figure}[h]
    \centering
    \fbox{\parbox{0.97\linewidth}{\footnotesize
    \begin{list}{}{%
        \setlength{\leftmargin}{2.4em}%
        \setlength{\itemindent}{-0.1em}%
        \setlength{\labelsep}{0.3em}%
        \setlength{\labelwidth}{2em}%
        \setlength{\itemsep}{0.5em}%
        \setlength{\parsep}{0pt}%
        \let\makelabel\relax
    }
    \item[{[30]}] Andrew et al. Luxton-Reilly. 2020. Novice Programming Mistakes: A Large-Scale Replication Study. In \emph{SIGCSE '20: Proceedings of the 51st ACM Technical Symposium on Computer Science Education}. ACM, 36--42.
    \item[{[56]}] Di Yang, Stuart MacNeil, and Andrew Petersen. 2024. Conversing with Copilot: Exploring Prompt Engineering for Solving CS1 Problems Using Natural Language. In \emph{Proceedings of the 2024 SIGCSE Technical Symposium on Computer Science Education}. ACM.
    \end{list}
    }}
    \caption{Examples of two ``hallucinated'' references in a paper from a SIGCSE in-cooperation venue (exactly as they appear in the PDF hosted on the ACM Digital library).}
    \label{fig:hallucinations}
    \Description{Two examples of hallucinated references}
\end{figure}

% \begin{figure}[h]
%     \centering
%     \includegraphics[width=1.0\linewidth]{images/hallucinate2.png}
%     \caption{Screenshots of two likely ``hallucinated references'' in a paper from a SIGCSE in-cooperation venue (exactly as they appear in the ACM Digital library)}
%     \label{fig:hallucinations}
%     \Description{Two examples of hallucinated references}
% \end{figure}

However, references are not always reliable. Citation errors have always existed, from minor mistakes such as incorrect years or errors in author names, to more serious cases where a reference is either
%incomplete, 
wrongly attributed, or cannot be matched to any real publication.  Two examples of this 
latter
%more serious 
type are shown in Figure \ref{fig:hallucinations}.  
These are reproduced verbatim from 
a 2025 paper published at a SIGCSE in-cooperation venue.
%a 2025 SIGCSE in-cooperation paper on the ACM Digital Library.
%These are copied verbatim as they appear on the ACM Digital Library in a paper published at a SIGCSE in-cooperation venue in 2025.  
In the first example in the figure (reference [30]) \emph{``et al.''} appears between the author's first and surname, there is no article beginning on page 36 of the SIGCSE 2020 proceedings, and (at least at the time of writing this article) the title \emph{``Novice Programming Mistakes: A Large-Scale Replication Study''} does not exist in the literature.  In the second example (reference [56]), while a paper with this title was published in 2023 (not 2024), it was written by three authors different from the ones listed.  Moreover, we could find no record of the three listed authors publishing together, and the middle author's name appears to be an erroneous variant of a well-known scholar in the community.

%As just one example of this latter type, a paper published in 2025 in a SIGCSE in-cooperation venue and hosted on the ACM Digital Library includes in its reference list three authors followed by the verbatim paper information: \emph{``2024. Conversing with Copilot: Exploring Prompt Engineering for Solving CS1 Problems Using Natural Language. In Proceedings of the 2024 SIGCSE Technical Symposium on Computer Science Education. ACM''}.  While a paper with this title was published in 2023, not 2024, it was written by three different authors than the ones listed.  Moreover, the three authors listed have never published together.  
One explanation for such mistakes is the growing use of large language models (LLMs) for academic writing support. LLMs are known to generate ``hallucinations'' which are plausible-looking outputs that are not grounded in real sources. This is especially problematic for references, where LLMs may fabricate citations by combining realistic author names, titles, venues, and publication metadata \cite{agrawal2024doLLMsknow}. Recent evidence suggests that such references are becoming more visible in the published literature \cite{yin2026llm, naddaf2026natureproblem}.  Within computing education, Denny, Becker, Leinonen, and Prather warned in a 2023 ITiCSE keynote that hallucinated references would begin appearing \cite{denny2023chat}, and later work found that LLM users are frustrated by hallucinated references during writing support \cite{zhou2024ethics}. Beyond computing education, Sch\"{o}ning reported encountering many such references while reviewing for a large SIGCHI conference, increasing the burden on peer review \cite{schoning2026chi30k}.

Computing education is an important field in which to study this issue. It has a long publication history within the ACM Digital Library, and a set of well-established venues that are sponsored by or in-cooperation with ACM SIGCSE, one of the oldest and largest of the ACM special interest groups. To our knowledge, there has been no large-scale analysis of reference integrity across the computing education literature spanning both pre- and post-LLM eras.  To address this gap, we conduct an analysis using a unique dataset: the entire ACM Digital Library, with publications spanning 1951-2026.  We explore publishing and referencing trends over this timeframe (to establish the scale of the issue), classify common bibliographic errors (to illustrate the challenge of automated detection), and we identify published references that appear to be LLM-hallucinations containing verifiably false information.  We organize the work around the following research questions:

%In this work, we analyse a unique dataset: the entire ACM Digital Library.  We identify computing education venues, examine referencing trends over time, classify several kinds of bibliographic errors, and flag references that are likely to be hallucinated.

\begin{description}
    \item[RQ1:] How have publication volumes and reference-list lengths in ACM computing education venues changed over time, and how do these trends compare with the broader ACM corpus?
    \item[RQ2:] What types of bibliographic errors appear in computing education references, and how common are these error types among references selected for manual review?
    \item[RQ3:] How many erroneous references in the computing education literature contain verifiably fabricated bibliographic information characteristic of LLM-generated hallucinations? %How often do invalid references in the computing education literature appear to be hallucinated?
\end{description}

% \begin{description}
%     \item[RQ1:] How have publication volumes and referencing practices in ACM computing education venues changed over time, and how do these trends compare with other ACM domains?
%     \item[RQ2:] What types of bibliographic errors occur in references in the computing education literature, and how frequently do they appear?
%     \item[RQ3:] How many erroneous references in the computing education literature contain verifiably fabricated bibliographic information characteristic of LLM-generated hallucinations? %How often do invalid references in the computing education literature appear to be hallucinated?
% \end{description}

%Our results show that hallucinated references are on the rise in published work within computing education. From this evidence, we argue that reference integrity must become a community-level responsibility for computing education. Authors, reviewers, program chairs, editors, and publishers must share responsibility for protecting research integrity from submission to publication. Finally, we present policy recommendations for the community to enact in support and of this position.

\section{Related Work}
The computing education community has a well-established practice of self-reflection, with numerous research studies using bibliometric analysis to explore and describe the nature of our community.  Citation analyses of ACE~\cite{lister2008ace,simon2020ace}, ICER~\cite{simon2016icer} ITiCSE~\cite{simon2020iticseauthors,simon2020iticsepapers}, and SIGCSE TS~\cite{lister2008citation}, keyword analysis of topics at ITiCSE and ICER~\cite{papamitsiou2020computing}, and later scientific collaboration network analysis~\cite{zhang2021scientific}, all provide evidence of a growing 
%computing education 
community, a broad range of topics, and strong international collaboration.  %Papamitsiou et al.~\cite{papamitsiou2020computing} used automated keyword analysis to describe topic evolution in computing education.  
Similar 
bibliometric 
approaches have been used for artificial intelligence literacy~\cite{tenorio2023artifical}, and automated assessment of programming education~\cite{paiva2023bibliometric}.

However, this prior research %involving bibliometric analysis in computing education 
is primarily descriptive in nature, and reflections on the \emph{quality} of scholarly work are less prevalent.  Notably, Simon et al.~\cite{simon2021confirmation} raised concerns about the integrity of our academic publications in computing education after an analysis of citations found evidence that authors did not always accurately represent the conclusions of papers they cited.
%studied the accuracy of citations involving introductory programming pass rates.   They found substantial evidence that authors did not always accurately represent the conclusions of the papers they are citing.
%, and expressed concerns about the rigour of our scholarly publication processes. 
%Concerns about the integrity of our scholarly publication processes have been recently reignited with the emergence of hallucinated references.
The emergence of hallucinated references has reignited such concerns about the rigor of our scholarly publication processes. %Prior work has examined this issue through automated detection methods, large-scale prevalence estimates, domain-specific audits, and broader discussions of the consequences for peer review and the scholarly record. 
%We organise the remainder of the related work around three themes: methods for detecting or estimating hallucinated references, the scholarly domains that have been studied, and evidence that the problem is increasing.

\subsubsection*{Detecting and Estimating Hallucinated References}

Humans are fallible and errors occasionally occur in otherwise sound work, which complicates distinguishing hallucinated references from ordinary bibliographic errors. %A central difficulty in studying hallucinated references is that they must be distinguished from ordinary bibliographic errors. %As a result, prior work has often relied on automated matching, fuzzy comparison, database lookup, or population-level estimation. 
Zhao et al. analysed 111 million references from 2.5 million papers across arXiv, bioRxiv, SSRN, and PubMed Central, estimating hallucination prevalence by comparing pre-LLM baseline unmatched-reference error rates with post-LLM rates~\cite{zhao2026llm}.  %rather than reference-level determinations. 
Similarly, Topaz et al.~\cite{topaz2026lancetfabricated} audited more than 125 million structured references in biomedical papers using automated checks against PubMed, Crossref, OpenAlex, and Google Scholar. %, followed by filtering and validation to identify fabricated citations .
These approaches are powerful for indicating large-scale changes, but provide only broad estimates that are not directly attributable to hallucinations without very careful manual verification.

Other work has proposed explicit verification systems. Agrawal et al.~\cite{agrawal2024doLLMsknow} and Chelli et al.~\cite{chelli2024hallucination} study the detection of hallucinated references using LLMs  under experimental conditions. 
%under experimental prompting conditions. %rather than references appearing in published papers 
Xu et al.~\cite{xu2026ghostcite} introduced CITEVERIFIER, which %parses references, queries bibliographic and web sources, and 
classifies candidate invalid citations using similarity matching, and Li et al. propose CITETRACER, a %multi-agent 
framework that combines %reference extraction,%PDF and BibTeX extraction, 
%evidence retrieval, 
deterministic field matching, and LLM-based adjudication for ambiguous cases \cite{li2026sourcedidnthappenmultiagent}. %Importantly, CITETRACER also provides a fine-grained taxonomy distinguishing real, potentially valid, and hallucinated citations, including field-level errors in titles, authors, venues, years, identifiers, and other metadata.

The closest methodological comparison to our work is Sakai et al.'s study of hallucinated references in ACL, NAACL, and EMNLP papers \cite{sakai2026hallucitation}. They combine large-scale citation extraction, fuzzy title matching against bibliographic databases, and manual verification of candidate hallucinated references. %Our work follows a similar principle of using automated processing to narrow the search space, but differs in focusing on computing education and using manual inspection to classify suspect references in a defined scholarly community.  %Together, this literature shows that 

\subsubsection*{A Growing Research-Integrity Problem}

Recent work suggests that hallucinated references are becoming more common. Zhao et al. estimate a sharp rise in non-existent references after widespread LLM adoption, including an estimated 146,932 hallucinated references in 2025 alone \cite{zhao2026llm}. Topaz et al. report that fabricated references in the biomedical literature increased substantially from 2023 to early 2026 \cite{topaz2026lancetfabricated}, and there  have been similar trends reported across AI/ML and security venues \cite{xu2026ghostcite}. Sakai et al. find a marked increase in NLP conference papers containing hallucinated references, from 20 papers in 2024 to 275 in 2025, commenting that they ``negatively affect the credibility of conferences'' \cite{sakai2026hallucitation}.

%This growth is concerning because hallucinated references can be difficult to detect, and currently requires manual verification~\cite{naddaf2026natureproblem}. [BETTER IN DISCUSSION]

%. They may contain plausible author names, realistic titles, credible venues, or partially correct metadata.

%Our work extends this literature by examining reference integrity within computing education. By combining ACM Digital Library metadata with manual inspection of suspect references, we treat hallucinated citations not only as a technical detection problem, but as a community-level issue for authors, reviewers, program chairs, editors, and publishers.

\section{Methods}

% Figure~\ref{fig:method} provides a high-level overview of our workflow.  

% , and we now detail the key steps. 
%which we briefly summarize now, before detailing the key steps.  We parsed ACM Digital Library metadata XML files into structured publication records (\textbf{Pubs}) and plain-text reference entries (\textbf{Refs}), identified computing education (CS Ed) venues, and extracted references from CS Ed publications published between 2021 and 2026 (\textbf{CS Ed Refs}).  We then parsed each reference into structured fields using a local LLM and attempted to match it to an ACM publication record using DOI and title information, producing matched references (\textbf{SET A}) and unmatched (\textbf{SET B}) references.  The full Pubs and Refs datasets were used to analyse publication and referencing trends (\textbf{RQ1}); suspicious matched references in SET A were manually coded to characterize common bibliographic errors (\textbf{RQ2}); and high-risk cases from Sets A and B were combined and manually verified to identify likely hallucinated references (\textbf{RQ3}).

\subsubsection*{ACM Digital Library data source}

With support of the ACM, on April 1st, 2026 we obtained a complete set of metadata for the digital library (ACM DL).  This was provided by ACM as a set of 28,269 .zip archives which, when extracted, resulted in 1,304,236 metadata XML files covering the years 1951 to 2026.  Most of these XML files contained the metadata for a single publication, such as authors, title, publication year, venue (e.g.\ book or conference proceedings), and/or identifiers (e.g.\ DOI), whilst some XML files represented non-publications (e.g.\ journal catalogues or conference proceedings lists).  XML files that contained metadata for a publication also listed entries in the reference list of that publication as plain text.  
%Each plain text reference was enclosed in \texttt{<ref/>} tags.

%The data used in this study were extracted from 1,304,236 ACM publication metadata XML files contained in 28,269 ZIP archives (of which 23,607 XML files in 2 ZIP archives were from the pilot data dump). 

\subsubsection*{Publication and Reference Data Processing}

\begin{figure}[ht]
    \centering
    \includegraphics[width=1.0\linewidth]{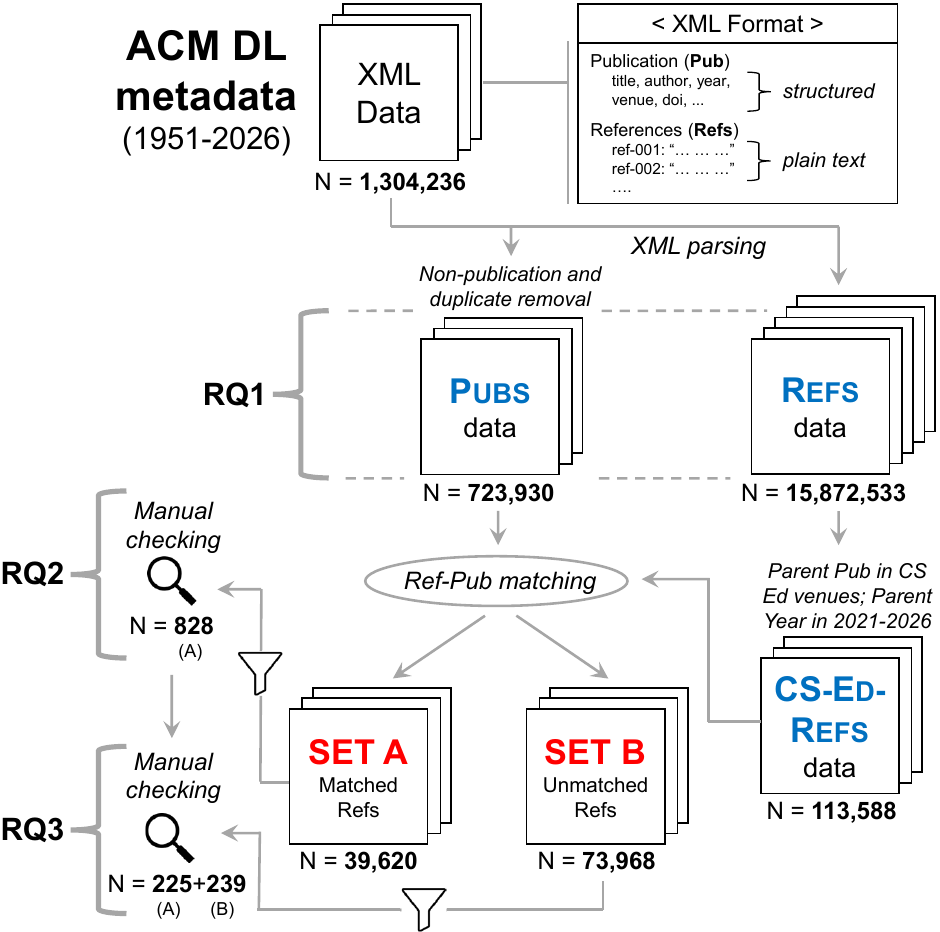}
    \caption{%High-level overview of method: 
    All publication (\textsc{Pubs}) and reference (\textsc{Refs}) data are parsed, a subset of reference data from CS Ed venues from 2021-2026 is extracted, and for every reference in this set, we attempt to find its corresponding structured entry in the publication data (using DOI and title).  This `Ref-Pub matching' step partitions the reference data into SET A (references where the DOI or title matches a corresponding entry in the \textsc{Pubs} dataset) and SET B (references without a match).}
    \label{fig:method}
    \Description{High-level overview of method}
\end{figure}

Figure~\ref{fig:method} provides a high-level overview of our workflow. 
%Figure \ref{fig:method} provides a high-level illustration of our data processing pipeline.  
We wrote a Python script to extract publication metadata (stored in structured fields) and reference data (stored as plain text) from these 1,304,236 XML files. 
%The extraction logic is described in Algorithm~\ref{alg:parse_zip}. 
This produced two datasets: \textsc{\textbf{Pubs}} for publications and \textsc{\textbf{Refs}} for references, where for each reference entry we also stored a link to its parent publication (as the unique path of the XML file). We removed non-publication entries 
%(whose XML files contain none of the author, title, year, or venue fields) 
and duplicate publications (along with their associated citations). The final \textsc{Pubs} dataset contained 723,930 ACM publication metadata entries, and the final \textsc{Refs} dataset comprised 15,872,533 plain-text reference entries.

% % ── Table 1 ─────────────────────────────────────
% \begin{table}[!h]
% \centering

% \caption{Summary of XML files processed and publications (Pubs) and references (Refs) dataset entries}
% \label{XML_counts}

% \begin{tabular}{ll}
% \toprule
% Item & Entry count \\
% \midrule
% XML files processed & 1,304,236 \\
% Publication (Pubs) dataset entries & 723,930 \\
% Reference (Refs) entries & 15,872,533 \\
% \midrule
% CS Ed publications (Pubs) entries & 24,751 \\
% CS Ed (2021--2026) (Pubs) entries & 5,225 \\
% CS Ed (2021--2026) (Refs) entries & 113,588\\
% \bottomrule

% \end{tabular}
% \end{table}
% % ── Table 1 end───────────────────────────────────

\subsubsection*{CS Ed venue identification}

We consulted authoritative works \cite{malmi2025topical, apiola2023venues} that have listed dedicated computing education venues to curate a set of ``CS Ed'' venues for identifying relevant work. 
%inclusion in our ``CS Ed'' dataset \cite{malmi2025topical, apiola2023venues}.  
We manually identified all variations of the following CS Ed venues in the \textsc{Pubs} dataset, and used their combined full venue names and acronyms to tag every entry in the \textsc{Pubs} dataset that was related to CS Ed:
%entry for inclusion in the CS Ed dataset:
%Given that acronyms are an unreliable way to uniquely identify a venue (e.g. the `ACE' acronym is used by multiple different conferences, including the Australasian Computing Education conference) we first generated a list of all unique venue names in the entire dataset.  Then, from that list, we manually tagged all of unique venue names that matched our curated list of computing education venues.  For space reasons we list just the acronyms here: 
ACE, CSERC, CompEd, ICER, ITICSE, Koli Calling, SIGCSE Bulletin, SIGCSE TS, SIGCSE Virtual, TOCE, UKICER, WCCCE, WiPSCE.  For CompEd, ITiCSE and SIGCSE Virtual we also included Working Group reports.  
%We also included variations of venue names (such as `ACSW' for the years that the ACE proceedings were published under the ACSW banner, and the `Baltic Sea conference' as an earlier name for Koli Calling). 
%Using the publication dataset, we manually examined all unique publication venue names and their acronyms and produced a mapping list to identify key ACM Computing Education (CSEd) venues across all year variations. 

\subsubsection*{\textsc{CS-Ed-Refs} dataset}
%Once all of the CS Ed venues were manually identified, we were able to automatically tag every \textsc{Pubs} entry that qualified as being in a ``CS Ed'' venue ($n$ = 24,751).  For every such \textsc{Pubs} entry, 
For every entry in \textsc{Pubs} that was CS Ed related ($n$ = 24,751), we then tagged all of its \textsc{Refs} entries.
%to indicate they were published at a CS Ed venue. 
%whether or not the reference appeared in a paper that was published at a CS Ed venue.
%We then produced a third dataset, 
From these, we produced a third dataset, 
\textsc{\textbf{CS-Ed-Refs}}, which consisted of all references that were from publications in CS Ed venues published since 2021 as of April 2026 (a total of 5,225 publications). We chose this year range to include several years prior to the widespread adoption of AI tools, whilst maintaining a feasible manual checking workload.  The \textsc{CS-Ed-Refs} dataset consisted of \textbf{113,588} reference entries.  

\subsubsection*{Reference-publication (Ref-Pub) auto-matching}

To support manual checking, we partitioned the \textsc{\textbf{CS-Ed-Refs}} dataset into two sets: \textbf{SET A} (Matched Refs) and \textbf{SET B} (Unmatched Refs). Because references in the raw ACM XML data were stored as \emph{highly varied} plain text, we first parsed each reference into structured fields---authors, title, venue, year, and DOI---using a locally hosted small LLM (Qwen3.5-4B~\cite{qwen35blog}) in deterministic, non-thinking mode. These parsed fields were used only for automated matching; manual checking referred back to the raw plain-text reference.

For each reference in the \textsc{\textbf{CS-Ed-Refs}} dataset, we applied automatic multi-tier fuzzy-matching to automatically identify a corresponding publication record in the complete \textsc{\textbf{Pubs}} dataset: 1)~matching by normalised DOI where available; 2)~otherwise, matching the normalized reference title against three normalized publication title variants (full title, main title, subtitle); 3)~failing that, performing a boundary-aware title-substring search to find the publication sharing the most consecutive significant words ($\geq$4 characters, excluding stopwords). 

%Tiers~2 and~3 applied a $\pm1$-year proximity gate when both publication years were available, to avoid matching a record with a similar title from a different year (but accepting adjacent years to include off-by-one year errors).

% For each reference, we attempted to identify a corresponding publication record in the complete \textbf{Pubs} dataset. We first matched references by normalized DOI (URL prefixes stripped, lowercased, trailing punctuation removed), where available. For references without a DOI match, we used title-based matching against ACM publication titles normalised for Unicode, whitespace, hyphenation, case, diacritics, punctuation.
% %\footnote{NFKC Unicode normalisation, exotic whitespace replaced with ASCII spaces, hyphen-break repair, lowercased, diacritics stripped (combining marks removed), all punctuation removed, runs of multiple spaces collapsed to single spaces. Trailing ``(abstract only)'' was also stripped. Applied to both the parsed title and publication titles before comparison.}, including full titles, main titles, and subtitles.  
% Title-based matching allowed close matches rather than requiring exact string identity, but were accepted only when they also satisfied a $\pm1$ year-proximity check where year information was available.

References linked automatically to an ACM publication record following this process were assigned to SET A; all remaining references were assigned to SET B. Thus, SET A contained 
\textbf{39,620} 
references that 
our automated matching procedure could link to an ACM metadata record,
%could be (fuzzy-likelihood )matched to ACM publication metadata, while 
while SET B contained \textbf{73,968}
references for which no corresponding ACM publication record was found.

\begin{figure*}[!htb]
  \centering
  \includegraphics[width=\linewidth]{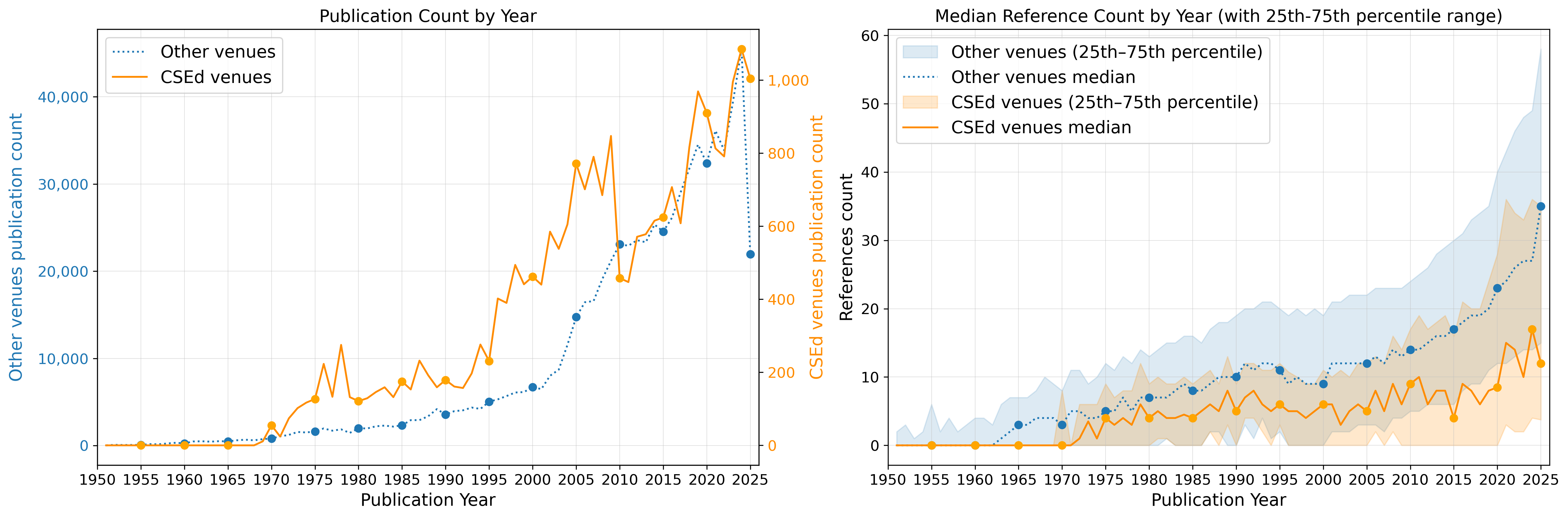}
  \caption{Number of publications per year (left) and median reference list length (right) from CSEd venues (solid line, right y-axis) and other venues (dotted line, left y-axis) over time (2026 data is incomplete for the year and thus excluded). %Each subplot uses a global y-axis for the count range.   \textbf{Left plot:} Count of unique publications per year for each venue category, plotted on dual y-axes to accommodate the different data range between categories. \textbf{Right plot:} Median number of references cited per publication per year, with shaded bands showing the interquartile range ($25^{th}$–$75^{th}$ percentile) for each venue category. 
}
  \label{fig:results_RQ1_trends}
  \Description{Publication and reference counts}
\end{figure*}

\subsection{Publication and Referencing Trends (RQ1)}

The problem of fabricated references is exacerbated by growing publication volumes and changing referencing practices.   To investigate publication and referencing trends over time, we compare publication and referencing rates in the computing education literature with those in other ACM domains.  
We analyzed the complete \textsc{Pubs} and \textsc{Refs} datasets, as well as the \textsc{Pubs} and \textsc{Refs} data when restricted only to CS Ed venues.  To visualize trends, we plotted: (1) the number of publications, and (2) the median length of reference lists in those publications, per year (1951-2026). 

\subsection{Common Bibliographic Errors (RQ2)}

To make manual inspection of \textbf{SET A} feasible, we focused on matched references whose author information appeared inconsistent with the corresponding ACM publication record. 
%SET A contained 39,620 matched reference-publication pairs, linked to ACM publications by DOI or title.
We identified candidate author discrepancies using an automated author-overlap check. The pipeline compared the (parsed) reference authors with the matched ACM author list, using exact surname matching with a fuzzy fallback, nearby given names or initials for confirmation, and special handling for truncated lists such as \emph{et al}. References with too little author overlap were flagged for manual checking, producing 1,278 candidate records.  We then manually screened these candidates to remove false-positive matches, i.e. records where the reference had been linked to an ACM publication with a similar-sounding title, but actually cited a URL-only source, such as a blog post or news article, or a non-ACM publication. Because such references did not correspond to their matched ACM record, author comparison against the ACM metadata was not meaningful. 
%would not provide a meaningful bibliographic-error classification. 
This final screening step left \textbf{828} records for detailed manual coding.  %Results of the classification of these are in Table \ref{tab:RQ2_reference_classification}.

Two coders divided the 828 records evenly within each year and classified references using an adapted version of Li et al.'s citation hallucination taxonomy~\cite{li2026sourcedidnthappenmultiagent}.  Where possible, we used the same letter codes, such as \texttt{H} for common `hallucination' errors (such as author and venue errors) and \texttt{R} for `real' references but with acceptable minor formatting variations.
%The error types observed in our data included: author-field errors, ACM metadata mismatches, valid bibliographic variants, Taskforce/Working Group/curriculum-document variations, alternative \emph{et al.} substitutions, and other typographical or human errors.
%checked whether each reference and matched publication described the same work. They 
Coders inspected the raw reference, matched metadata, ACM DOI page, source PDF, and, where useful, external bibliographic records. Source PDFs were checked carefully before labelling cases as potentially hallucinated (i.e. with code \texttt{H}), since some apparent errors reflected ACM metadata discrepancies rather than errors in the published paper.

%Records were 

\subsection{Hallucinated References (RQ3)}

For the purposes of RQ3, we use a deliberately conservative definition of a ``hallucinated'' reference. We classify a reference as hallucinated only when it contains \emph{verifiably fabricated bibliographic
information}.  Incomplete references are not treated as hallucinations when the information that is present is correct and the cited work can be identified. For example, references with omitted authors or missing venue information are not classified as hallucinated unless they also contain fabricated or clearly false bibliographic content.

To identify references that might meet this definition, we constructed a conservative candidate set from both matched and unmatched references. From \textbf{SET A}, we selected all references coded \texttt{H} during RQ2 manual coding, indicating a potential hallucination or serious author-field error. This produced \textbf{225} candidates, typically cases where the title linked to a
real ACM publication but the author information was inconsistent.

From \textbf{SET B}, we selected references that appeared to cite a CS Ed venue, producing 904 candidates. Because these references had not been automatically linked to ACM publications, we further checked each parsed title using the Semantic Scholar \texttt{/paper/search/match} endpoint. This identified \textbf{239} entries for which no title match was returned. We treated these only as candidates, since missing matches may reflect indexing gaps, title-parsing errors, or unusual formatting rather than fabrication.

The final RQ3 candidate set therefore combines Set A entries with inconsistent core metadata, especially authorship, and Set B entries whose titles could not be matched automatically to known publications. 
For each of these 464 candidates, we manually searched for a record online.  Entries with a credible publication record were removed; those without one were checked by a second reviewer using bibliographic databases, publisher pages, DOI searches, and general web search.  We classified a reference as hallucinated only when careful and repeated manual searches failed to identify a
corresponding valid publication record and when the reference contained verifiably fabricated bibliographic information. 

\section{Results}

\subsection{Publication and Referencing Trends (RQ1)}

Figure~\ref{fig:results_RQ1_trends} shows that publication volume in CS Ed venues has grown substantially over time, particularly from the mid-1990s onwards. The visible dip around 2010 is due to the fact that until 2009, SIGCSE Technical Symposium records appeared as both conference proceedings entries and as entries in the \emph{SIGCSE Bulletin}, with distinct DOI and venue records in the ACM metadata.

Reference lists in CS Ed papers have also become longer over time. Although the increase is less pronounced than in the broader ACM corpus, the upward trend is clear.  As both publication volumes and reference-list lengths grow, the number of citations that authors, reviewers, and publishers must verify also increases.

% Figure~\ref{fig:results_RQ1_trends} shows publication and referencing trends for CS Ed venues compared with other ACM venues. The number of CS Ed publications has grown substantially over time, particularly from the mid-1990s onwards. The main exception is a visible dip around 2010. This appears to reflect a change in how SIGCSE Technical Symposium papers were represented in the ACM Digital Library rather than a contraction in computing education publishing. Until 2009, SIGCSE Technical Symposium papers appeared both as conference proceedings entries and as entries in \emph{SIGCSE Bulletin}, with distinct ACM records. Because these records were associated with distinct DOIs and venues, they were not initially removed by our duplicate-checking process. 

% Figure~\ref{fig:results_RQ1_trends} also shows that reference lists in CS Ed papers have become longer over time. Although the median number of references per paper has not increased as sharply as in the broader ACM corpus, the upward trend is still clear. This makes reference integrity increasingly important.  As both publication volumes and reference-list lengths grow, the number of citations that authors, reviewers, and publishers must verify also increases. 

% \begin{quote}
% \small
% Amber Arnold, Tracy Camp, Wendy DuBow, and Ivory Johnson. 2023. K-12 Teacher Experiences from Online Professional Development for Teaching APCSA. In \textit{Proceedings of the 54th ACM Technical Symposium on Computer Science Education V. 1}. ACM, 974--980.
% \end{quote}

\begin{table}[ht]
\caption{Reference classification codes and counts.}
\label{tab:RQ2_reference_classification}
\centering
\small
\begin{tabular}{@{}l r p{0.66\columnwidth}@{}}
\toprule
Code & Count & Category descriptor \\
\midrule
S & 229 & ACM metadata mismatch; PDF shows correct information, with no evidence of hallucination \\
H & 225 & Author-field error, including addition, deletion, reordering, or potential fabrication \\
R & 188 & Real reference; valid bibliographic variant \\
W & 150 & Working Group, Task Force, and ACM/IEEE reports \\
%W & 149 & Working Group or minor character/format variation, such as dash or ampersand changes \\
M & 22 & Other typo or apparent human error \\
A & 14 & Alternative substitution for \textit{et al.}, such as ``and al'' or ellipses \\
\midrule
Total & 828 & \\
\bottomrule
\end{tabular}
\end{table}

\subsection{Common Bibliographic Errors (RQ2)}

Table~\ref{tab:RQ2_reference_classification} summarises the outcome of the manual coding process for data in SET A. Surprisingly, the most common issue was an apparent mismatch between the ACM DL XML reference data and the plain-text reference as printed in the camera-ready PDF (code \texttt{S}, 229 cases). A frequent example involved \emph{et al.}: in some records, the plain-text reference extracted from the ACM XML data omitted \emph{et al.} or otherwise altered the author list, while the PDF itself contained a clean and correct reference. We therefore treat these as metadata mismatches rather than author errors in the published paper.

As title data had already been matched for SET A entries, the next largest category was author-field errors, which are classified as ``hallucinated'' in the taxonomy of Li et al. \cite{li2026sourcedidnthappenmultiagent} (code \texttt{H}, 225 cases).  Deletions were by far the most common kind of author-related error, where one or more authors were missing from the reference, sometimes from the middle of an author list.  One such example is shown below, where the correct author list should be \emph{Krueger, Huang, Liu, Santander, Weimer and Leach}:

\begin{sidequote}
Ryan Krueger, Tyler Santander, Westley Weimer, and Kevin Leach. 2020. Neurological divide: An fMRI study of prose and code writing. In Proceedings of the 42nd International Conference on Software Engineering (ICSE’20). IEEE/ACM.
\end{sidequote}

%These were cases where the title generally referred to a real publication, but the author list in the reference did not match the corresponding record. Deletions were especially common, with one or more authors missing from the reference, sometimes from the middle of an author list. 

%If all other author information was correct, and there were no other errors in the reference, we did not treat such deletions as potential hallucinations. 

Several other categories reflected valid or non-hallucinatory variation. 
%Non-ACM references (code N, 216 cases) included sources such as websites, blogs, and other materials outside the ACM publication dataset. 
Valid bibliographic variants (code \texttt{R}, 188 cases) referred to real publications where the reference differed from the ACM record in an acceptable way. Working Group reports, Joint Task Force curriculum reports, and similar ACM materials were also cited in many different formats (code \texttt{W}, 150 cases), sometimes with incomplete or omitted author lists, but with otherwise valid information. Two typical examples of this, referring to the same report, are below:

\begin{sidequote}
CC Taskforce, Computing Curricula 2020: Paradigms for Global Computing Education November 2020.

\medskip
Clear, A., Impagliazzo, J., Parrish, A, and Wang, P. 2020. Computing Curricula 2020: Paradigms for Global Computing Education. DOI: 10.1145/3467967, https://dl.acm.org/citation.cfm?id=3467967
\end{sidequote}

% \begin{quote}
% \small
% CC Taskforce, Computing Curricula 2020: Paradigms for Global Computing Education November 2020.
% \end{quote}

% \begin{quote}
% \small
% Clear, A., Impagliazzo, J., Parrish, A, and Wang, P. 2020. Computing Curricula 2020: Paradigms for Global Computing Education. DOI: 10.1145/3467967, https://dl.acm.org/citation.cfm ?id=3467967
% \end{quote}

%Finally, a small number of records involved alternative substitutions for \emph{et al.} (code A, 14 cases) or other %minor typographical and likely human errors (code M, 22 cases).

A small number of records involved minor typos as well as acceptable formatting variations (codes \texttt{M} and \texttt{A}).

\begin{table}[t]
\caption{Hallucinated reference data by venue. Total refers to the number of hallucinated references, and Papers refers to the number of papers containing at least one.}
\label{tab:RQ3_hallucinations}
\centering
\small
\begin{tabular}{@{}l r l r r@{}}
\toprule
Venue & Year & Volume & Total & Papers \\
\midrule
SIGCSE Technical Symposium & 2026 & Volume 1 & 13 & 4 \\
SIGCSE Technical Symposium & 2026 & Volume 2 &  4 & 3 \\
SIGCSE Technical Symposium & 2025 & Volume 1 &  1 & 1 \\
SIGCSE Technical Symposium & 2025 & Volume 2 &  2 & 2 \\
UKICER    & 2025 & --       &  7 & 1 \\
CompEd    & 2025 & Volume 2 &  1 & 1 \\
TOCE      & 2025 & --       &  1 & 1 \\
ITiCSE    & 2025 & Volume 1 &  1 & 1 \\
\midrule
Total     &      &          & 30 & 14 \\
\bottomrule
\end{tabular}
\end{table}

\subsection{Hallucinated References (RQ3)}

Table~\ref{tab:RQ3_hallucinations} summarizes references we classified as ``hallucinated'' (containing verifiably fabricated bibliographic
information) after conservative manual verification. 
%We ignored reference errors where there was no fabricated content (e.g. author deletions but without other errors). 
From the 464 candidates,
%In total, 
we identified 30 hallucinated references appearing across 14 papers. 
%These were references for which repeated manual searches could not find a corresponding publication record.  
In most cases, there was more than one clue that suggested a reference was hallucinated.  
In the example below, the title of the paper corresponds to a real publication (with correct venue), however \emph{all authors are incorrect}.  Moreover, the page reference shown \emph{is invalid}, as no paper starts on page 974 of the SIGCSE 2023 proceedings:
%Consider this example:

\begin{sidequote}
\small
Amber Arnold, Tracy Camp, Wendy DuBow, and Ivory Johnson. 2023. K-12 Teacher Experiences from Online Professional Development for Teaching APCSA. In Proceedings of the 54th ACM Technical Symposium on Computer Science Education V. 1. ACM, 974-980.
\end{sidequote}

In a similar example below, the title of the paper again corresponds to a real publication (with correct venue and DOI). Although the authors are all prominent scholars in the field of computing education, they did not author this paper, nor any papers together in 2023. 

\begin{sidequote}
\small
David Weintrop, Shriram Krishnamurthi, and Armando Fox. 2023. Programming Is Hard -- Or at Least It Used to Be: Educational Opportunities And Challenges of AI Code Generation. In Proceedings of the 54th ACM Technical Symposium on Computer Science Education (SIGCSE). doi:10.1145/ 3545945.3569759
\end{sidequote}

The (incorrect) authors listed in both of the examples shown are likely unaware of these misattributions.  Examples such as these that we could verify as hallucinated appeared only in the most recent part of the dataset. We found hallucinated references in 2025 papers from five CS Ed venues: SIGCSE TS, UKICER, CompEd, TOCE, and ITiCSE. The largest number outside SIGCSE TS occurred in a single UKICER 2025 paper, which contained seven hallucinated references.  One example of these is shown below, where we could find no record of the title \emph{``Exploring the Motivational Role of LLMs in Programming Education''} via any online search, and it certainly does not appear in the SIGCSE 2025 proceedings:

\begin{sidequote}
\small
Andrzej Boguslawski et al. 2025. Exploring the Motivational Role of LLMs in Programming Education. In Proceedings of the 2025 SIGCSE Technical Symposium on Computer Science Education. ACM.
\end{sidequote}

%In this example, we could find no record of the title \emph{``Exploring the Motivational Role of LLMs in Programming Education''} via any online search, suggesting it was entirely fabricated.   

%In this example, we could find no record of the title \emph{``Exploring the Motivational Role of LLMs in Programming Education''} via any online search, suggesting it was entirely fabricated.   
%and it does not appear in the SIGCSE 2025 proceedings

The 2026 data must be interpreted carefully because our ACM DL snapshot was obtained on April 1, 2026, and therefore only captures publications available in the first part of the year. In particular, UKICER, ITiCSE, and CompEd had not yet run in 2026. For the purposes of comparison between years, the SIGCSE TS is useful as both the 2025 and 2026 proceedings were present in our dataset. In SIGCSE TS, the number of verified hallucinated references increased from 3 in the 2025 proceedings to 17 in the 2026 proceedings. These 17 references appeared across seven SIGCSE TS 2026 publications (including four full papers in Volume 1).

Overall, these results suggest that hallucinated references remain relatively rare in absolute terms, but they are no longer isolated anomalies. They have appeared across multiple computing education venues, and the SIGCSE TS comparison indicates a sharp increase between 2025 and 2026. Given that our identification process % was intentionally 
used intentionally restrictive criteria,
%very conservative, 
these counts should be interpreted as a conservative lower-bound estimate of the problem.

\section{Discussion}

Our findings show that hallucinated references are no longer a hypothetical concern for computing education research. 
%While the absolute numbers we have reported remain small, 
%our 
Our results represent a conservative lower-bound estimate and provide evidence of a recent increase that mirrors trends being reported in other domains \cite{sakai2026hallucitation}. 

%although our deliberately conservative verification process means that these results should be interpreted as a lower-bound estimate rather a complete count. 
%this is nevertheless concerning, especially considering the sharp increase year on year. 
%This problem is difficult to address through ordinary peer review alone. 

%Many references that we initially
%flagged turned out not to be hallucinations, but ACM metadata
%mismatches, valid bibliographic variants, or likely human errors.
%Many references that we found initially suspicious turned out to not be hallucinations, but ACM metadata mismatches, valid variants, or most likely human errors. 
%Some invalid references are obvious once inspected closely, but many are superficially plausible, especially when they contain real author names and realistic titles. 

The hallucinated references we identified were not uniform. Thirteen of the 30 appeared to be entirely fabricated, with no corresponding publication record found. The remaining 17 were hybrid references, combining a real title with fabricated or incorrect authorship, venue, or year information.  In all 17 cases, at least some author names were fabricated, which aligns with recent experimental work on hallucinated citations showing that author fields fail far more often than other fields \cite{chen2026fake}.
In seven of these hybrid cases, at least one listed author was correct. 
%This variety in structure makes hallucinated references difficult to detect. 
This mixture of real and fabricated elements is what makes the problem difficult.  Many hallucinated references are not obviously fake, but plausible recombinations of details that may partially match real publications.

Both human and automated approaches to checking are therefore challenging. 
Many references that we found initially suspicious turned out not to be hallucinations, but ACM metadata mismatches, valid variants, or likely human errors. Some invalid references are obvious once inspected closely, but many are superficially plausible, especially when they contain real (or legally changed) author names and realistic titles. These cases often required substantial manual checking, which goes beyond what can be expected from reviewers. This problem is compounded by increasing publication volumes, longer reference lists, stretched reviewer capacity \cite{adam2025peer, horta2024crisis, vitak2024beyond}, and the growing use of generative AI in research workflows
\cite{rangarajan2026outpaced, schoning2026chi30k}.

%Automated checking is therefore attractive, but our results suggest that it is not straightforward. The most common issue in our RQ2 analysis was an apparent mismatch between ACM XML reference data and the reference as printed in the camera-ready PDF. In many of these cases, the PDF contained a clean reference, while the metadata record appeared incomplete or altered. Automated reference-checking systems can only be as good as publisher metadata allows.

Automated checking will likely become more common, but our results show that it is not straightforward. The most frequent issue in our RQ2 analysis was an apparent mismatch between ACM XML reference data and the reference as printed in the camera-ready PDF. In many cases, the PDF contained a clean reference, while the metadata record appeared incomplete or altered (special characters were often the cause, such as for one paper where the author ``C. I. Ches{\~n}evar'' appeared in the XML metadata as ``C. I. Ches nevar''). Automated reference-checking systems are therefore limited by the quality of the metadata they use as ground truth. 
At the same time, superficial checks may miss subtle hybrid hallucinations.
For these reasons, tools should be designed as human-in-the-loop systems that highlight suspicious references for inspection, explain why a reference has been flagged, and return feedback to authors before papers are sent out for review. The error categories we identify in this work provide a starting point for such systems.

%These cases often required substantial manual checking, which goes beyond what can be expected from reviewers. This problem is compounded by increasing publication volumes, longer reference lists, stretched reviewer capacity, and the growing use of generative AI in research workflows \cite{rangarajan2026outpaced, schoning2026chi30k, vitak2024beyond}. % As the number of papers and references increases, so does the verification burden placed on authors, reviewers, program chairs, editors, and publishers. These pressures make it unrealistic to assume that reviewers alone can reliably detect every fabricated or unverifiable reference.  Clearly technical solutions to this problem will be on the horizon, and can be used to check articles submitted for review \cite{abbonato2026checkifexist}. 

While our current results are a cause for concern, the problem is likely even larger at the submission stage. Our analysis only examined the published versions of papers that had already passed peer review and publication checks. Given typical conference and journal acceptance rates, 
the published literature is likely only the visible part of a much larger submission-stage problem.
%and assuming that accepted papers are on average more carefully prepared than rejected submissions, it is reasonable to assume that more papers containing hallucinated references have been submitted than published. 
A special session by the program chairs of the 2025 SIGCSE TS brought some of these concerns to the community \cite{sigcse2025proceedings}.

%James note: I don't think this paragraph is necessary. It's clear we aren't naming and shaming anyone.
%Our purpose in reporting these cases is not to identify or criticise individual authors, but to provide an empirical account of an emerging publication integrity problem. For this reason, we do not provide links to the papers containing hallucinated references. Instead, we report aggregate counts by venue and year, and use selected examples only to illustrate the kinds of reference errors observed. Hallucinated references may arise through different mechanisms, including careless AI-assisted writing, reference-manager errors, or even copying references from other sources such as student work.

The wider scholarly community is beginning to respond. Naddaf and Quill describe hallucinated citations as an emerging research-integrity problem across scholarly publishing \cite{naddaf2026natureproblem}, and Resnik and Hosseini argue that hallucinated citations may constitute research misconduct 
%when citations function as data in scholarly papers 
\cite{resnik2026accountability}. 
Publishers and venues are also developing more explicit policies. ACM policy states that content-integrity issues arising from the use of AI during authorship may lead to retraction; arXiv has proposed lengthy bans for authors who submit fabricated references \cite{chawla2026researchers}; and some computing education venues
%(including TOCE and SIGCSE TS) 
now list fabricated references among the grounds for desk rejection\footnote{\url{https://dl.acm.org/journal/toce/author-guidelines}}. 
Naddaf and Quill also report a hard-line response in which authors submitting fabricated references to one journal are prohibited from resubmitting the work \cite{naddaf2026natureproblem}.% These examples show that hallucinated references are increasingly being treated not as minor formatting mistakes, but as publication-integrity failures.

%James note: I think the below paragraph is larely redundant and can be cut. I have commented it out for now.
%The harder question is what should happen after hallucinated references are found in already-published work. There is likely no single appropriate response. Some cases may be best handled through corrections, especially where the fabricated reference is not central to the argument and can be removed or replaced without changing the paper's claims. Other cases may require stronger action, including retraction, if fabricated references support substantive claims, misrepresent prior work, or undermine the evidential basis of the paper. In our opinion, the response should be proportionate to the role played by the fabricated reference in the paper, while still recognising that any presence of hallucinated references weakens trust in the paper and scholarship.

Based on our findings, we propose a shared responsibility model as the way forward. Authors should verify every cited work, especially when generative AI has been used during writing, and should not cite sources that they have not personally checked.  
%This view is aligned with that of the SIGCSE TS 2027 program chairs, who reminded all authors prior to the submission deadline that they will \emph{``not prescreen for hallucinated references as we believe this is every researcher's responsibility''}.   
Reviewers and program committees should not be expected to manually audit every reference, but venues can adopt targeted checks for high-risk cases and make policies explicit at submission. Publishers 
should work to 
%also have an important role to play. Our results showed that metadata problems can make valid references appear suspicious, so publishers should look to 
improve metadata quality and reduce discrepancies between publication records and camera-ready PDFs, as well as support post-publication correction, and provide tools for reference validation.

\section{Conclusions}

References are essential to scholarly work, allowing claims to be checked, credit to be assigned, and evidence to accumulate. 
In this paper, we examined reference integrity across the ACM computing education literature, combining large-scale metadata analysis with conservative manual verification. 
Most suspicious references were not hallucinations, but metadata mismatches, valid variants, or ordinary citation irregularities.
%Most suspicious references were not hallucinations, but metadata mismatches, valid variants, non-ACM sources, or ordinary citation irregularities. 
However, we also identified 30 
hallucinated references containing fabricated information,
%verified hallucinated references 
appearing across 14 papers in 2025 and 2026.  Most of these appeared in the SIGCSE TS 2026 proceedings, where 2.3\% of papers contained at least one hallucinated reference. 
%, including an increase at SIGCSE TS from 3 such references in 2025 to 17 in 2026. Because our process was intentionally conservative, these counts should be read as a lower bound. 
The problem is still small, but it is real, recent, and growing. This is an important issue for the computing education field, and it requires the shared responsibility of authors, reviewers, program chairs, editors, and publishers. %Addressing the problem of hallucinated references right now is crucial when it is still manageable.

\begin{acks}
This work was supported by the Research Council of Finland grant \#356114. 
\end{acks}

\balance
\bibliographystyle{ACM-Reference-Format}
\bibliography{references}

\end{document}